\documentclass[journal]{IEEEtran}

\usepackage[T1]{fontenc}
\usepackage{graphicx}
\usepackage{cite}
\usepackage{subcaption}

\usepackage{overpic}
\usepackage{amssymb}
\usepackage{amsmath}
\usepackage{amsthm}
\usepackage{array}
\usepackage{color}
\usepackage{hyperref}

\IEEEoverridecommandlockouts

\theoremstyle{plain}

\newtheorem{lemma}{Lemma}
\newtheorem{corollary}{Corollary}
\newtheorem{remark}{Remark}
\newtheorem{proposition}{Proposition}

\newcommand{\vect}[1]{\mathbf{#1}}

\def\diag{\mathrm{diag}}

\def\Ttran{\mbox{\tiny $\mathrm{T}$}}
\def\CN{\mathcal{N}_{\mathbb{C}}} %Complex Gaussian
\def\Rbar{\bar{R}}
\def\hbsr{\vect{h}_{\mathrm{sr}}}
\def\hsd{{h}_{\mathrm{sd}}}
\def\hsr{h_{\mathrm{sr}}}
\def\hrd{h_{\mathrm{rd}}}
\def\hbrd{\vect{h}_{\mathrm{rd}}}
\def\betasr{\beta_{\mathrm{sr}}}
\def\betard{\beta_{\mathrm{rd}}}
\def\betasd{\beta_{\mathrm{sd}}}
\def\betaIRS{\beta_{\mathrm{IRS}}}
\def\Gt{G_{\mathrm{t}}}
\def\Gr{G_{\mathrm{r}}}

\begin{document}

\title{\huge Intelligent Reflecting Surface vs.~Decode-and-Forward: \\ How Large Surfaces Are Needed to Beat Relaying?}

\author{
\IEEEauthorblockN{Emil Bj{\"o}rnson, \emph{Senior Member, IEEE}, {\"O}zgecan {\"O}zdogan, \emph{Student Member, IEEE}, Erik G. Larsson, \emph{Fellow, IEEE}
\thanks{ \vspace{-3mm}
\newline\indent \copyright 2019 IEEE. Personal use of this material is permitted. Permission from IEEE must be obtained for all other uses, in any current or future media, including reprinting/republishing this material for advertising or promotional purposes, creating new collective works, for resale or redistribution to servers or lists, or reuse of any copyrighted component of this work in other works. This version corrects a typo in Proposition 3, which also affected Figure 5. Please refer to \href{https://doi.org/10.1109/LWC.2024.3453787}{10.1109/LWC.2024.3453787} for additional clarification.
\newline\indent The paper was supported by ELLIIT and the Swedish Research Council. 
\newline \indent The authors are with the Department of Electrical Engineering (ISY), Link\"{o}ping University, SE-58183 Link\"{o}ping, Sweden \{emil.bjornson,ozgecan.ozdogan,erik.g.larsson\}@liu.se.}
\vspace{-6mm}
}}

\maketitle

\begin{abstract}
The rate and energy efficiency of wireless channels can be improved by deploying software-controlled metasurfaces to reflect signals from the source to the destination, especially when the direct path is weak.
While previous works mainly optimized the reflections, this letter compares the new technology with classic decode-and-forward (DF) relaying. The main observation is that very high rates and/or large metasurfaces are needed to outperform DF relaying, both in terms of minimizing the total transmit power and maximizing the energy efficiency, which also includes the dissipation in the transceiver hardware. \vspace{-1mm}
\end{abstract}

\begin{IEEEkeywords}
Intelligent reflecting surface, DF relaying. \vspace{-3mm}
\end{IEEEkeywords}

\vspace{-2mm}

\IEEEpeerreviewmaketitle

\section{Introduction}

A reflectarray is a surface that ``reflects'' an impinging plane wave in the shape of a beam \cite{Huang2005a}. Different from parabolic reflectors, whose physical curvature and direction determine the beamforming, a reflectarray is flat and consists of an array of discrete elements that each scatter and phase-shift the impinging waves differently \cite{Pozar1997a}. The phase-shift pattern among the elements determine in which direction the reflected beam is formed. While the surface can be large, the individual elements are typically sub-wavelength sized \cite{Headland2017,Alu2016}. Reflectarrays with real-time reconfigurable properties have recently gained interest in mobile communications, under names such as \emph{intelligent reflecting surface (IRS)} \cite{Wu2018a,Wu2019a} and \emph{software-controlled metasurfaces} \cite{Liaskos2018a,Bjornson2019d,Renzo2019a}. The main idea is to support the transmission from a source to a destination by adapting the propagation environment; that is, to configure the IRS to beamform its received signal towards the destination.

This is the same use case as for half-duplex relays \cite{Laneman2004a}, with the key difference that a relay actively processes the received signal before retransmitting an amplified signal, while an IRS passively reflects the signal without amplification but with beamforming. The relay achieves a higher signal-to-noise ratio (SNR) at the cost of a pre-log penalty due to the two-hop transmission. A comparison with an ideal amplify-and-forward (AF) relay was made in \cite{Huang2018a}, showing large energy efficiency gains by using an IRSs. However, decode-and-forward (DF) relaying is known to outperform AF relaying in terms of achievable rates \cite{Farhadi2009a} and is thus a better benchmark.

In this letter, we try to make a fair comparison between IRS-supported transmission and repetition-coded DF relaying, with the purpose of determining how large an IRS needs to be to outperform conventional relaying. To this end, we optimize both technologies by computing the optimal transmit powers and the optimal number of elements in an IRS.

\vspace{-3mm}

\section{System Model}
\label{sec:system-model}

We consider communication from a single-antenna source to a single-antenna destination. The deterministic flat-fading channel is denoted by $\hsd \in \mathbb{C}$. The received signal at the destination is
\begin{equation}
y = \hsd  \sqrt{p} s + n,
\end{equation}
where $p$ is the transmit power, $s$ is the unit-power information signal, and $n \sim \CN(0,\sigma^2)$ is the receiver noise. For notational convenience, the antenna gains are included in the channels.
The capacity of this single-input single-output (SISO) channel is 
\begin{align} 
R_\mathrm{SISO} =  \log_2 \left( 1 + \frac{p|\hsd|^2}{\sigma^2}  \right).  \label{eq:R_SISO}
\end{align}
The capacity can potentially be increased by involving additional equipment in the communication. In this paper, we consider two such setups: An IRS that is configured to ``reflect'' the signal towards the destination or a relay that operates in DF mode. The respective achievable rates (also known as spectral efficiencies) are derived below and then optimized analytically to enable a fair comparison. However, we stress that the choice of system model is biased in favor of the IRS; in particular, the assumption of deterministic flat-fading channels is ideal for an IRS since it is less capable of handling channel estimation and frequency-selective fading than relays.

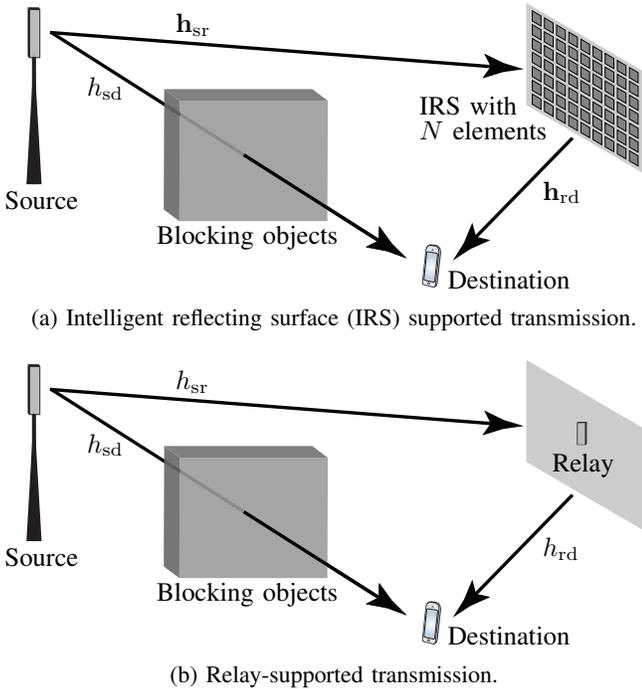
\begin{figure} 
        \centering 
        \begin{subfigure}[b]{\columnwidth} \centering 
	\begin{overpic}[width=.95\columnwidth,tics=10]{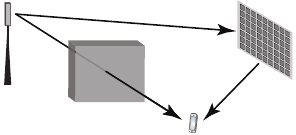}
	\put(-2,13){Source}
	\put(68,0.5){Destination}
	\put(22,7.5){Blocking objects}
	\put(63.5,28){IRS with }
	\put(63.5,24){$N$ elements}
	\put(11,31){$\hsd$}
	\put(25,41){$\hbsr$}
	\put(83,14.5){$\hbrd$}
\end{overpic} 
                \caption{Intelligent reflecting surface (IRS) supported transmission.} \vspace{2mm}
                \label{fig:IRSexample}
        \end{subfigure} 
        \begin{subfigure}[b]{\columnwidth} \centering  \vspace{+2mm}
	\begin{overpic}[width=.95\columnwidth,tics=10]{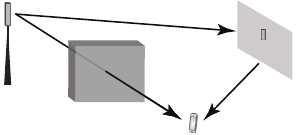}
	\put(-2,13){Source}
	\put(68,0.5){Destination}
	\put(22,7.5){Blocking objects}
	\put(84.5,28){Relay}
	\put(11,31){$\hsd$}
	\put(25,41){$\hsr$}
	\put(83,14.5){$\hrd$}
\end{overpic}  
                \caption{Relay-supported transmission.}  
                \label{fig:DFexample} 
        \end{subfigure} 
        \caption{Illustration of the two setups considered in this paper.} 
        \label{fig:examples}  \vspace{-2mm}
\end{figure}

\vspace{-3mm}

\subsection{IRS-supported Transmission}

In this setup, we have an IRS with $N$ discrete elements, as illustrated in Fig.~\ref{fig:examples}(a). The deterministic channel from the source to the IRS is denoted by $\hbsr \in \mathbb{C}^N$, where $[\hbsr ]_n$ denotes the $n$th component. The channel between the IRS and the destination is denoted by $\hbrd \in \mathbb{C}^N$.
Each element has a smaller size than the wavelength, thus it scatters the incoming signal with approximately constant gain in all directions of interest \cite{Ozdogan2019a}.\footnote{Experiments with the element size $\lambda/5 \times \lambda/5$ is presented in \cite{Headland2017} and   $\lambda/8 \times \lambda/8$ is considered in \cite{Alu2016}, where $\lambda$ denotes the wavelength.} The IRS's properties are therefore fully represented by the diagonal matrix
\begin{equation}
\vect{\Theta} = \alpha \diag\left( e^{j \theta_1}, \ldots, e^{j \theta_N} \right),
\end{equation}
where $\alpha \in (0,1]$ is the fixed amplitude reflection coefficient and $\theta_1,\ldots,\theta_N$ are the phase-shift variables that can be optimized by the IRS. Following the system model derived in \cite{Ozdogan2019a} (which was earlier used in e.g. \cite{Wu2018a,Huang2018a,Nadeem2019a}), the received signal at the destination is
\begin{equation}
y = (\hsd + \hbsr^{\Ttran} \vect{\Theta} \hbrd) \sqrt{p} s + n,
\end{equation}
where $p$, $s$, and $n$ are defined as in the SISO case. 
Since the channels are deterministic, the destination knows them perfectly and the phase-shift variables can be optimized.\footnote{Channel estimation is non-trivial for IRS-supported transmissions, but some recent methods are found in \cite{He2019a}.}

\begin{lemma}
The channel capacity of the IRS-supported network is
\begin{align} \label{eq:R_IRS_prel}
R_\mathrm{IRS}(N) &= \max_{\theta_1,\ldots,\theta_N} \log_2 \left(
1 + \frac{p  | \hsd + \hbsr^{\Ttran} \vect{\Theta} \hbrd|^2}{\sigma^2} \right) \\ 
& =  \log_2 \left( 1 + \frac{p(|\hsd|+ \alpha \sum_{n=1}^{N} \! | [\hbsr]_n \, [\hbrd]_n | )^2}{\sigma^2}  \right) \!.  \label{eq:R_IRS}
\end{align} 
\end{lemma}
\begin{IEEEproof}
For any given $\vect{\Theta}$, the rate expression in \eqref{eq:R_IRS_prel} is achieved from the capacity of an additive white Gaussian noise channel. Notice that $\hbsr^{\Ttran} \vect{\Theta} \hbrd  = \alpha \sum_{n=1}^{N} e^{j \theta_n}[\hbsr]_n [\hbrd]_n$. The maximum rate, which is the capacity, is achieved when the phase-shifts are selected as $\theta_n = \arg(\hsd) - \arg( [\hbsr]_n [\hbrd]_n)$ to give every term in the sum the same phase as $\hsd$.\footnote{This proof idea follows the same main steps as in \cite[Sec.~III.B]{Wu2018a}.}
\end{IEEEproof}

\subsection{Relay-supported Transmission}

In this alternative setup, we make use of a half-duplex relay that is deployed at the same location as the IRS. This setup is illustrated in Fig.~\ref{fig:examples}(b). We consider the classic repetition-coded DF relaying protocol where the transmission is divided into two equal-sized phases. In the first phase, the source transmits and the received signal at the destination is
\begin{equation} \label{eq:y1d}
y_{1\mathrm{d}} = \hsd \sqrt{p_1} s + n_{1\mathrm{d}},
\end{equation}
where $p_1$ is the transmit power, $s$ is the unit-power information signal, and $n_{1\mathrm{d}} \sim \CN(0,\sigma^2)$ is the receiver noise.
In the same phase, the received signal at the relay is
\begin{equation}
y_{1\mathrm{r}} = \hsr \sqrt{p_1}s + n_{1\mathrm{r}},
\end{equation}
where $\hsr \in \mathbb{C}$ denotes the channel between the source and relay, while $n_{1\mathrm{r}} \sim \CN(0,\sigma^2)$ is the receiver noise. The DF relay uses $y_{1\mathrm{r}}$ to decode the information and then encodes it again for transmission in the second phase. Note that the relay can be compact; an antenna, transceiver chains, and a baseband unit fit into the dimensions of a small mobile phone.

In the second phase, the relay transmits $\sqrt{p_2} s$ and the received signal at the destination is
\begin{equation} \label{eq:y2d}
y_{2d} = \hrd \sqrt{p_2} s + n_{2\mathrm{d}},
\end{equation}
where $p_2$ is the transmit power, $\hrd \in \mathbb{C}$ denotes the channel between the relay and destination, while $n_{2\mathrm{d}} \sim \CN(0,\sigma^2)$ is the receiver noise. By utilizing \eqref{eq:y1d} and \eqref{eq:y2d} for maximum ratio combining, the following rate is achievable at the destination.

\begin{lemma}
The achievable rate with repetition-coded DF relaying is
\begin{equation} \label{eq:R_DF}
R_\mathrm{DF} = \frac{1}{2} \log_2 \left( 1 + \min \left(  \frac{p_1 |\hsr|^2}{\sigma^2} ,   \frac{p_1|\hsd|^2 }{\sigma^2}  +  \frac{p_2 |\hrd|^2}{\sigma^2}  \right)  \right).
\end{equation}
\end{lemma}
\begin{IEEEproof}
This is a classical result found in \cite[Eq.~(15)]{Laneman2004a}.
\end{IEEEproof}

\begin{remark}
For brevity, the analysis in this letter assumes deterministic channels, but the extension to fading channels with perfect channel knowledge is straightforward: we only need to take expectations of the rate expressions in \eqref{eq:R_IRS} and \eqref{eq:R_DF}. Hence, all the conclusions apply to this case as well.
\end{remark}

\section{Analytical Performance Comparison}
\label{sec:analytical-comparison}

In this section, we compare the three achievable rates that were presented in Section~\ref{sec:system-model}. 
Interestingly, the expressions only depend on the amplitudes of the channel elements, but not on their phases. 
For brevity, we introduce the notation $| \hsd |= \sqrt{\betasd}$, $|\hsr| = \sqrt{\betasr}$, $|\hrd| = \sqrt{\betard}$, and $\frac{1}{N} \sum_{n=1}^{N} \! | [\hbsr]_n \, [\hbrd]_n | = \sqrt{\betaIRS} $.
We can now rewrite \eqref{eq:R_SISO}, \eqref{eq:R_IRS}, and \eqref{eq:R_DF} in the more compact forms
\begin{align} 
R_\mathrm{SISO} &=  \log_2 \left( 1 + \frac{p \betasd}{\sigma^2}  \right),  \label{eq:R_SISO_2} \\
\label{eq:R_IRS_2}
R_\mathrm{IRS} (N) &= \log_2 \left( 1 + \frac{p(\sqrt{\betasd}+N \alpha \sqrt{\betaIRS} )^2}{\sigma^2}  \right), \\
R_\mathrm{DF} &= \frac{1}{2} \log_2 \left( 1 +  \min \! \left(  \frac{p_1 \betasr}{\sigma^2} ,  \frac{p_1 \betasd}{\sigma^2}  +   \frac{p_2 \betard}{\sigma^2} \right) \! \right) \!. \label{eq:R_DF_2}
\end{align}
It is obvious that $R_\mathrm{IRS} (N) \geq R_\mathrm{SISO}$ since equality is achieved for $N=0$ and $R_\mathrm{IRS}(N) $ is an increasing function of $N$. In fact, the rate grows as $\mathcal{O}(\log_2(N^2))$ when $N$ is large, as previously noted in \cite{Wu2018a} and further explained in \cite{Bjornson2019e}.
The comparison between the IRS and DF relay cases is non-trivial. To make it fair, we first select $p_1$ and $p_2$ optimally, while having the same average power $p$ as when using the IRS.

\begin{proposition} \label{proposition:DF-power-allocation}
Assume that $p_1,p_2\geq 0$ are selected under the constraint $p = \frac{p_1+p_2}{2}$.
If $\betasd>\betasr$, it holds that $R_\mathrm{SISO} > R_\mathrm{DF} $ for any selection of $p_1,p_2$, thus DF relaying is suboptimal.

If $\betasd \leq \betasr$, the rate with DF relaying is maximized by $p_1 = \frac{2p \betard}{\betasr+\betard-\betasd}$ and $p_2 = \frac{2p (\betasr-\betasd)}{\betasr+\betard-\betasd}$, leading to
\begin{equation} \label{eq:R_DF_3}
R_\mathrm{DF} = \frac{1}{2} \log_2 \left( 1 +  \frac{2p \betard \betasr}{(\betasr+\betard-\betasd) \sigma^2} \right).
\end{equation}
\end{proposition}
\begin{IEEEproof}
If $\betasd>\betasr$, $\min \! \left(  \frac{p_1 \betasr}{\sigma^2} ,  \frac{p_1 \betasd}{\sigma^2}  +   \frac{p_2 \betard}{\sigma^2} \right)= \frac{p_1 \betasr}{\sigma^2}$, which is maximized by $p_1 = 2p$ and $p_2=0$. Hence, the relay is not used and obviously $R_\mathrm{SISO}>R_\mathrm{DF} $.
If $\betasd \leq \betasr$, $R_\mathrm{DF} $ is maximized by selecting $p_1,p_2$ to achieve $\frac{p_1 \betasr}{\sigma^2} = \frac{p_1 \betasd}{\sigma^2}  +   \frac{p_2 \betard}{\sigma^2}$ under the constraint $p = \frac{p_1+p_2}{2}$. This gives a linear system of equations with the solution that is stated in the proposition.
\end{IEEEproof}

One important implication of Proposition~\ref{proposition:DF-power-allocation} is that the relay-supported network needs to switch between two modes: SISO transmission and DF relaying. It is only when the channel from the source to the relay is stronger than the direct path from the source to destination (i.e., $\betasr \geq \betasd$) that DF relaying might provide $R_\mathrm{DF} > R_\mathrm{SISO}$.

\begin{proposition} \label{prob-IRS-vs-DF}
The IRS-supported transmission provides the highest rate for any $N \geq 1$ if $\betasd>\betasr$. In the case $\betasd \leq \betasr$, it provides the highest rate if and only if
\begin{align} \label{eq:minimum-N}
N >  \frac{\sqrt{ \left( \sqrt{1 + \frac{2p \betard \betasr}{(\betasr+\betard-\betasd) \sigma^2}  }- 1 \right) \frac{\sigma^2}{p} } - \sqrt{\betasd}}{\alpha \sqrt{\betaIRS}}.
\end{align}
\end{proposition}
\begin{IEEEproof}
Since $R_\mathrm{IRS} (N) > R_\mathrm{SISO}$ for $N\geq 1$, the IRS-supported case gives the highest rate if and only if $R_\mathrm{IRS} (N) > R_\mathrm{DF} $. This always occurs for $\betasd>\betasr$ since $ R_\mathrm{SISO}>R_\mathrm{DF}$ due to Proposition~\ref{proposition:DF-power-allocation}. If $\betasd \leq \betasr$, the inequality $R_\mathrm{IRS} (N) > R_\mathrm{DF} $ can be simplified to \eqref{eq:minimum-N} by utilizing \eqref{eq:R_IRS_2} and \eqref{eq:R_DF_3}.
\end{IEEEproof}

To interpret the result in Proposition~\ref{prob-IRS-vs-DF}, we now consider that there is line-of-sight to and from the IRS. We assume that each IRS element has the same size as the relay's antenna, thus it follows that all elements in $\hbsr$ have the same magnitude as $\hsr$ and all elements in $\hbrd$ have the same magnitude as $\hbrd$. Consequently, $\betaIRS =\betasr \betard$. We first notice that
although IRS-supported transmission provides the highest rate for $\betasd>\betasr$, the difference between $R_\mathrm{IRS}(N)$ and  $R_\mathrm{SISO}$ is small in this case since $\sqrt{\betasd} \gg N \alpha \sqrt{\betaIRS} = N \alpha \sqrt{\betasr \betard}$ for most practical values of $N$ because $\betard$ is a very small number in practice; note that a ``large'' channel gain in wireless communications is $-60$\,dB. Hence, it is in the case $\betasd \leq \betasr$ that an IRS can provide an appreciable performance gain.

The right-hand side of \eqref{eq:minimum-N} depends on the transmit SNR $p/\sigma^2$, the amplitude reflection coefficient $\alpha$, and the channel gains $\betasd$, $\betasr$, and $\betard$ (recall that  $\betaIRS =\betasr \betard$). 
Note that the right-hand side approaches $-\frac{\sqrt{\betasd}}{\alpha \sqrt{\betasr \betard}}$ as $p \to \infty$, which implies that the IRS-supported transmission achieves the largest rate at high SNR for any $N$.
In contrast, the inequality in \eqref{eq:minimum-N} becomes
\begin{align} \label{eq:N-ineq-lowSNR}
N >  \frac{\sqrt{ \frac{1}{(\betasr+\betard-\betasd) }  } - \frac{\sqrt{\betasd}}{\sqrt{\betasr \betard}} }{\alpha }.
\end{align} 
as $p\to 0$, which can be a very large number if $\betasd \ll \betasr$. For example, \eqref{eq:N-ineq-lowSNR} becomes $N>963$ for
$\alpha=1$, $\betasd=-110$\,dB, $\betasr=-80$\,dB, and $\betard=-60$\,dB. 

In summary, the choice between an IRS and a relay depends on the SNR and number of elements. In Section~\ref{sec:numerical}, we assess if practical setups operate in the low or high SNR regime; that is, if any of the asymptotic results above can applied.

\subsection{Transmit Power Minimization Under Rate Constraints}

If the destination requires a particular data rate $\Rbar$, the rate expressions in \eqref{eq:R_SISO_2}--\eqref{eq:R_DF_3} can be used to identify the required transmit power for each of the three communication setups.

\begin{corollary}
To achieve a data rate $\Rbar$, the SISO case requires the power 
\begin{equation}
p_{\mathrm{SISO}}= \left(2^{\Rbar}-1 \right) \frac{\sigma^2}{\betasd},
\end{equation}
the IRS-supported transmission requires the power
\begin{equation}
p_{\mathrm{IRS}}(N) =  \left(2^{\Rbar}-1 \right) \frac{\sigma^2}{(\sqrt{\betasd}+N \alpha \sqrt{\betaIRS} )^2},
\end{equation}
and the relay-supported transmission requires the power
\begin{equation}
p_{\mathrm{DF}} =  \begin{cases}
 \left(2^{2\Rbar}-1 \right) \frac{\sigma^2}{\betasd} & \textrm{if } \betasd>\betasr, \\
 \left(2^{2\Rbar}-1 \right) \frac{(\betasr+\betard-\betasd) \sigma^2}{2 \betard \betasr} & \textrm{if } \betasd \leq \betasr. \\
\end{cases}
\end{equation}
\end{corollary}

If the relay-supported system switches between SISO and DF relaying mode to minimize the transmit power, its required transmit power is $p_{\mathrm{DFmode}} = \min(p_{\mathrm{SISO}},p_{\mathrm{DF}} )$.

\subsection{Total Power Minimization Under Rate Constraints} \label{subsec:total_power}

The total power consumption, $P_{\mathrm{total}}$, of the system consists of both transmit power and dissipation in hardware components. In the SISO case, it is $P_{\mathrm{total}}^{\mathrm{SISO}} =  p_{\mathrm{SISO}}/\nu + P_{\textrm{s}} + P_{\textrm{d}}$, where $\nu\in(0,1]$ is the efficiency of the power amplifier while $P_{\textrm{s}}$ and $P_{\textrm{d}}$ are the hardware-dissipated power at the source and destination, respectively. In the IRS case, it becomes \cite{Huang2018a}
\begin{equation} \label{eq:Ptot_IRS}
P_{\mathrm{total}}^{\mathrm{IRS}}(N) =  \frac{p_{\mathrm{IRS}}(N)}{\nu} + P_{\textrm{s}} + P_{\textrm{d}} + N P_{\textrm{e}},
\end{equation}
where $P_{\textrm{e}}$ is the power dissipation per element, which is caused by the circuitry required for adaptive phase-shifting. In the relaying case, the source is only active half of the time, thus
\begin{equation} \label{eq:Ptot_DF}
P_{\mathrm{total}}^{\mathrm{DF}} =  \frac{p_{\mathrm{DF}}}{\nu} + \frac{1}{2} P_{\textrm{s}} + P_{\textrm{d}} + P_{\textrm{r}} ,
\end{equation}
where $P_{\textrm{r}}$ is the hardware-dissipated power at the relay.

\begin{proposition} \label{prop:N-opt-ee}
Suppose $\betaIRS$ is a constant independent of $N$.
For a given data rate $\Rbar$, the total power $P_{\mathrm{total}}^{\mathrm{IRS}}(N)$ when using an IRS  is a convex function and minimized by \vspace{-2mm}
\begin{equation} \label{eq:N-opt-ee}
N^{\mathrm{opt}} = \sqrt[3]{ 
 \frac{ 2 \left(2^{\Rbar}-1 \right) \sigma^2}{\nu \alpha^2 \betaIRS P_{\textrm{e}}}
}
- \frac{1}{\alpha} \sqrt{ \frac{\betasd}{\betaIRS} }.
\end{equation}
\end{proposition}
\begin{IEEEproof}
The function is convex since $\frac{\partial^2}{\partial N^2} P_{\mathrm{total}}^{\mathrm{IRS}}(N)>0$. The solution \eqref{eq:N-opt-ee} is then obtained from $\frac{\partial}{\partial N} P_{\mathrm{total}}^{\mathrm{IRS}}(N)=0$.
\end{IEEEproof}

The optimal number of elements in \eqref{eq:N-opt-ee} is generally not an integer number, thus the true optimum is either the closest smaller or larger integer. The optimum can also be negative, making the SISO case with $N=0$ the true optimum.
One example when $\betaIRS$ is independent of $N$ is the line-of-sight case with $\betaIRS =\betasr \betard$ described in the previous subsection.\footnote{If
$[\hbsr]_n \, [\hbrd]_n$ for $n=1,\ldots,N$ are realizations from independent and identically distributed random variables, then $\betaIRS \to \mathbb{E}\{|[\hbsr]_n \, [\hbrd]_n| \}$ as $N \to \infty$ due the law of large numbers. The upper bound could equal $\sqrt{\betasr \betard}$ since the IRS is deployed at the same location as the relay.}

\section{Numerical Performance Comparison}
\label{sec:numerical}

\begin{figure}[t!]
	\centering  
	\begin{overpic}[width=.9\columnwidth,tics=10]{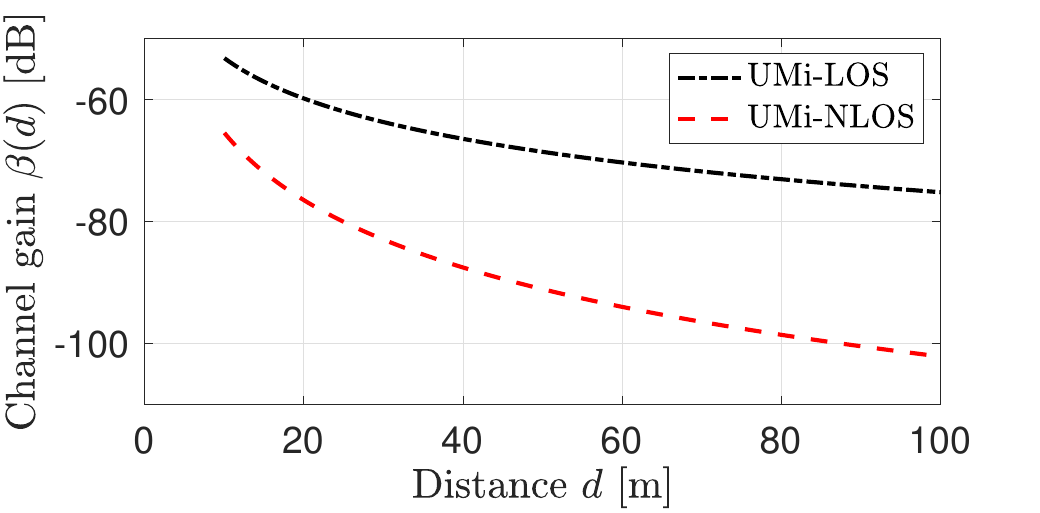}
\end{overpic} 
	\caption{Typical channel gains as a function of the distance, when including the antenna gains $\Gt=\Gr=5$\,dBi.}
	\label{figureChannelGain}
\end{figure}

We will now compare the systems numerically. The channel gains are modeled using the 3GPP Urban Micro (UMi) from \cite[Table B.1.2.1-1]{LTE2010b} with a carrier frequency of 3\,GHz. We use the line-of-sight (LOS) and non-LOS (NLOS) versions of UMi, which are defined for distances $\geq\!10$\,m. We let $\Gt$ and $\Gr$ denote the antenna gains (in dBi) at the transmitter and receiver, respectively.
We neglect shadow fading to get a deterministic model and show the channel gain $\beta$ as a function of the distance $d$ in Figure~\ref{figureChannelGain}:
\begin{align} \notag
&\beta(d) \,\mathrm{[dB]} \\ &= \Gt + \Gr +  \begin{cases}
-37.5 - 22 \log_{10}(d/1\,\textrm{m}) & \textrm{if LOS}, \\
-35.1 - 36.7 \log_{10}(d/1\,\textrm{m}) & \textrm{if NLOS}. \\
\end{cases}  \label{eq:pathloss}
\end{align}
We included Fig.~\ref{figureChannelGain} to reinforce the point that a seemingly small number such as $-60$\,dB is actually a \emph{very large} channel gain, while typical numbers are in the range $-70$ to $-110$\,dB.

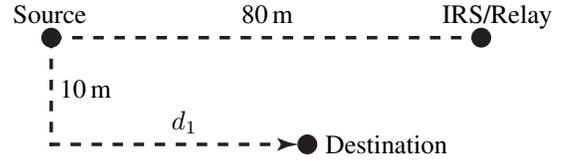
\begin{figure}[t!]
	\centering 
	\begin{overpic}[width=.7\columnwidth,tics=10]{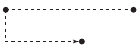}
	\put(-4,30){Source}
	\put(63,2){Destination}
	\put(88,30){IRS/Relay}
	\put(45,30){80\,m}
	\put(6,14){10\,m}
	\put(30,7){$d_1$}
\end{overpic} 
	\caption{The simulation setup where $d_1$ is a variable.}
	\label{figureSetup}  \vspace{-2mm}
\end{figure}

We consider the simulation setup in Fig.~\ref{figureSetup}, where the source and IRS/relay are deployed at fixed locations, while the location of the destination is determined by the variable $d_1$. We use \eqref{eq:pathloss} to compute the channel gains based on the distances, assuming equal-sized 5\,dBi antennas at the source and IRS/relay, while the destination is a handset with an omnidirectional antenna with 0\,dBi. The IRS and relay are deployed to have LOS channels to the source, and the destination has a LOS channel to the IRS/relay. We can therefore assume $\betaIRS =\betasr \betard$. Moreover, there is an NLOS channel between the source and destination, which leads to a weaker channel gain and motivates the use of an IRS or relay to support the transmission.

\begin{figure} 
        \centering 
        \begin{subfigure}[b]{\columnwidth} \centering 
	\begin{overpic}[width=.95\columnwidth,tics=10]{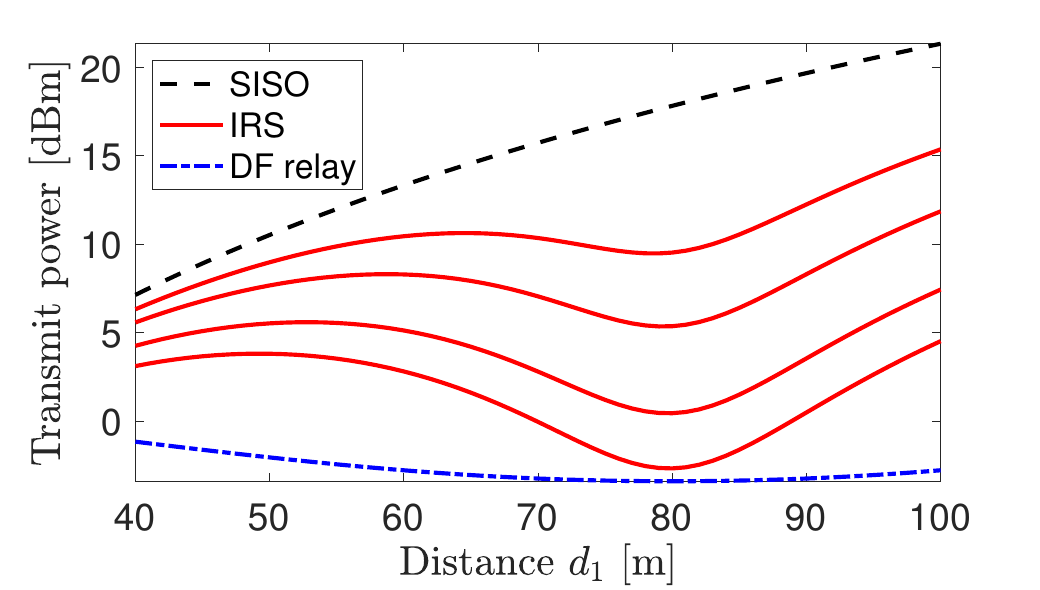}
	\put(13,15){$N=25,50,100,150$}
		\put(54,34.5){\vector(-1,-1){16}}
\end{overpic} 
                \caption{$\Rbar=4$\,bit/s/Hz.} 
                \label{fig:Powerexample1}
        \end{subfigure} 
        \begin{subfigure}[b]{\columnwidth} \centering 
	\begin{overpic}[width=.95\columnwidth,tics=10]{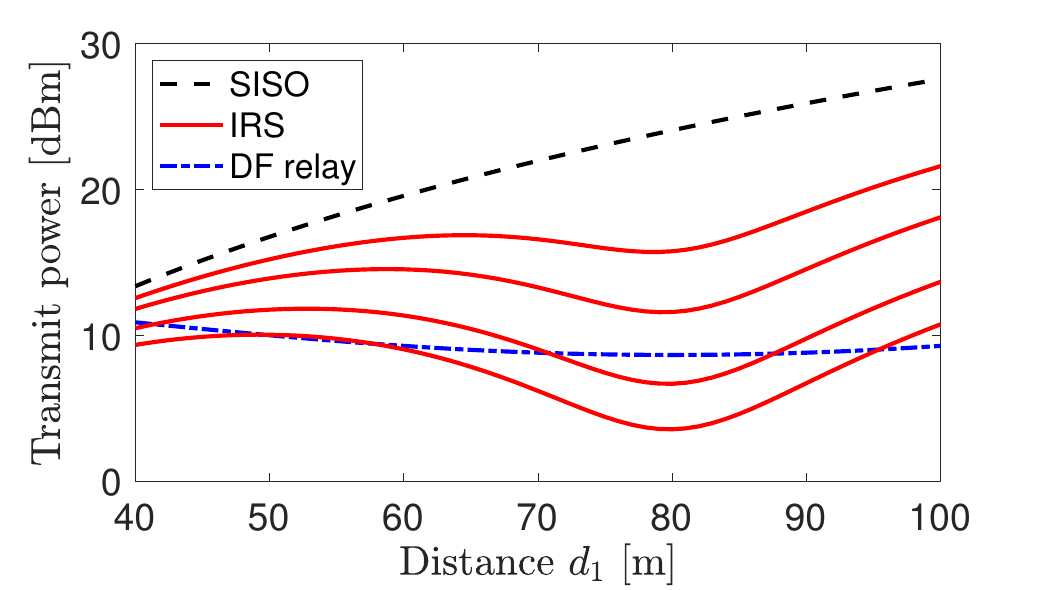}
	\put(16,14.5){$N=25,50,100,150$}
	\put(53,35.5){\vector(-1,-1){17}}
\end{overpic}  
                \caption{$\Rbar=6$\,bit/s/Hz.}  
                \label{fig:Powerexample2} 
        \end{subfigure} 
        \caption{The transmit power needed to achieve the rate $\Rbar$ in the scenario shown in Fig.~\ref{figureSetup}, as a function of the distance $d_1$.}  
        \label{fig:simulationPower}  \vspace{-2mm}
\end{figure}

Fig.~\ref{fig:simulationPower} shows the transmit power that is needed to achieve a rate of either $\Rbar=4$\,bit/s/Hz or $\Rbar=6$\,bit/s/Hz. The bandwidth is $B=10$\,MHz, the corresponding noise power is $-94$\,dBm, and $\alpha=1$. The figure shows results for the SISO case, DF relaying (without mode selection), and an IRS with $N\in \{ 25, \, 50, \, 100,\, 150\}$.
In the case of $\Rbar=4$\,bit/s/Hz, the SISO case requires the highest power while the DF relaying case requires the least power at all the considered locations of the destination. The transmit power required in the IRS case reduces as $N$ increases and the gap to the DF relaying case is smallest when the destination is either close to the source or to the IRS. By using \eqref{eq:minimum-N}, we obtain that $N>164$ is required if the IRS should outperform DF relaying when $d_1=80$\,m.

Higher transmit powers are needed in Fig.~\ref{fig:simulationPower}(b), where the rate is increased to $\Rbar=6$\,bit/s/Hz. The IRS case becomes more competitive; it requires the least power when the destination is close to the source, while ``only'' $N>76$ is needed to outperform relaying when $d_1=80$\,m. The reason that relaying loses some of its advantages is that it must have a higher SINR than in the IRS case due the $1/2$-prelog penalty; thus, the required power grows more rapidly with $\Rbar$ with relaying.

\subsection{Energy Efficiency}

\begin{figure}[t!]
	\centering  
	\begin{overpic}[width=.95\columnwidth,tics=10]{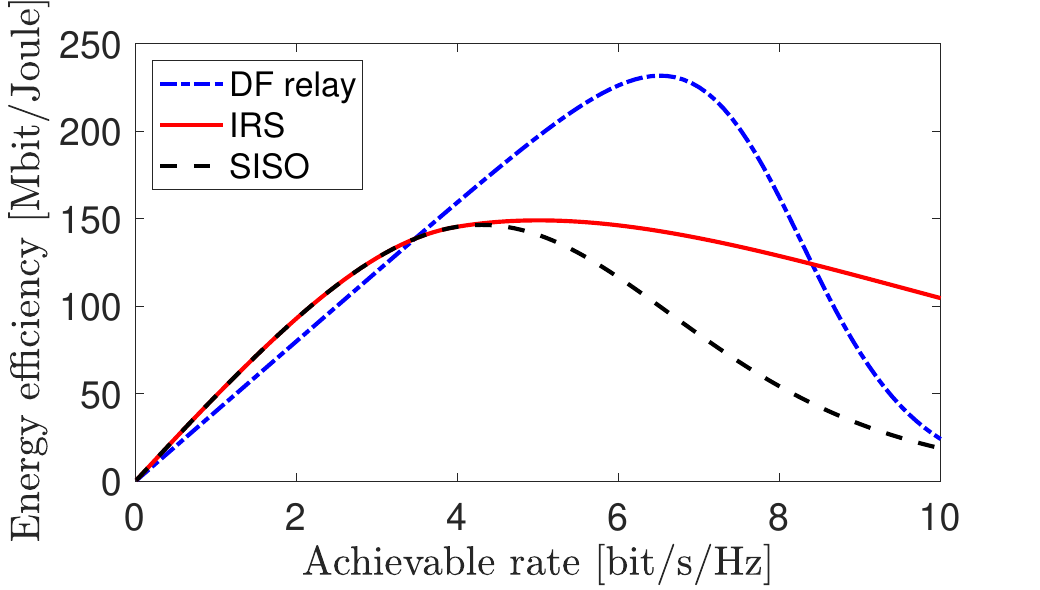}
\end{overpic} \vspace{-1mm}
	\caption{The energy efficiency as a function of the rate $\Rbar$.}
	\label{fig:simulationEE} \vspace{-3.5mm}
\end{figure}

It was shown in \cite{Huang2018a} that an IRS can improve the energy efficiency (EE), which is defined as $B \cdot \Rbar / P_{\mathrm{total}}$. We will make a similar analysis using the models in Sec.~\ref{subsec:total_power} with $\nu = 0.5$, $P_{\textrm{s}} = P_{\textrm{d}} = P_{\textrm{r}} = 100$\,mW, $P_{\textrm{e}} = 5$\,mW \cite{Huang2018a}, and $d_1=70$\,m. Fig.~\ref{fig:simulationEE} shows the EE as a function of $\Rbar$. The number of elements in the IRS, $N$, is optimized for maximal EE using Proposition~\ref{prop:N-opt-ee}. 
The SISO case provides the highest EE for $\Rbar \in (0,3.47]$\,bit/s/Hz, while the DF relaying case provides the highest EE for $\Rbar \in (3.47,8.41]$\,bit/s/Hz. It is only for $\Rbar>4$\,bit/s/Hz that the IRS has $N^{\mathrm{opt}}>0$ and it is only for $\Rbar>8.41$\,bit/s/Hz that it provides higher EE than DF relaying.
Hence, a system that switches between the SISO and DF relaying modes is preferable both in terms of minimizing the transmit power and maximizing the energy efficiency, except when very high rates are required.

\section{Conclusion and Discussion}

We have compared classic repetition-coded DF relaying with the new concept of IRSs. The key observation is that an IRS needs hundreds of reconfigurable elements (each of the size of an antenna) to be competitive---even if we considered ideal phase-shifting and frequency-flat channels, which are two assumptions that clearly benefit the IRS. The reason is that the source's transmit power must travel over two channels to reach the destination, leading to a very small channel gain $\betasr \betard$ per element in the IRS---the SNR becomes almost the same as for amplify-and-forward relaying \emph{without amplification}. Hence, the IRS needs many elements to compensate for the low channel gain. In contrast, with DF relaying, we first transmit over a channel with gain $\betasr$ and then transmit again over a channel with gain $\betard$. While the large number of elements is a weakness for IRSs, the advantage is that an IRS requires no power amplifiers in its ideal form; however, in practice, active components are needed for adaptive phase-shifting. Even if the power dissipation per element is low, the total power is non-negligible. An IRS only achieves higher EE than DF relaying if very high rates are needed.  Note that we only considered repetition-coded DF relaying, but other DF protocols achieve higher rates by optimizing the coding of the two hops and, thus, are even more competitive against an IRS-supported transmission \cite{Khormuji2009a}.

The fact that the source and destination are physically separated from the IRS is the key feature---it allows for controlling the propagation environment---but also the reason for the large pathlosses. Classical reflectarrays are using nearby sources equipped with high-gain horn antennas to manage the pathloss \cite{Pozar1997a}. Although the IRS will be larger than a relay, it is important to notice that an IRS with hundreds of elements, which was necessary to beat DF relaying in the simulations of this paper, can be still physically rather small since each element is assumed to have a sub-wavelength size  \cite{Headland2017,Alu2016}.
In general, it is the total size of the IRS (and not the number of elements or their individual size) that determines the pathloss, as explained in detail in \cite{Ozdogan2019a}. For the sizes considered in this paper, the IRS will not behave as a specular reflector but reflect the incident wave as a beam; however, there are other scenarios where that might occur, particularly when operating in the THz range.

\textit{\textbf{Reproducible research:}} The simulation code can be downloaded from \url{https://github.com/emilbjornson/IRS-relaying}

\bibliographystyle{IEEEtran}

\bibliography{IEEEabrv,refs}

\end{document}